\documentclass[useAMS,usenatbib]{mn2e}
\usepackage{graphicx}
\usepackage{lscape}
\usepackage{ulem}

\def\kms{km~s$^{-1}$}

\def\lya{Ly$\alpha$}

\def\hkpc{$h_{70}^{-1}$ kpc}
\def\dind{$D$-index}
\def\mg2{Mg~II}
\def\ebmv{E($B-V$)}
\def\bmk{$B-K$}

\title[Intervening absorbers in GRB060418]
{Three intervening galaxy absorbers towards GRB060418:  faint and dusty?}
\author[S. L. Ellison et al.]
{Sara L. Ellison$^1$\thanks{Email: sarae@uvic.ca},
Paul Vreeswijk$^{2,3}$,
C\'edric Ledoux$^3$,
Jon P. Willis$^1$,
\newauthor Andreas Jaunsen$^4$,
Ralph A. M. J. Wijers$^5$,
Alain Smette$^3$,
Johan P. U. Fynbo$^6$,
\newauthor Palle M\o ller$^7$,
Jens Hjorth$^6$,
Andreas Kaufer$^3$
\\
$^1$ Dept. Physics \& Astronomy, University of Victoria, 3800 Finnerty Rd, 
Victoria, BC, V8P 1A1, Canada\\
$^2$ Departamento de Astronom\'ia, Universidad de Chile, Casilla 36-D,
  Santiago, Chile\\
$^3$ European Southern Observatory, Alonso de C\'ordova 3107, Casilla
  19001, Santiago, Chile\\
$^4$ Institute of Astrophysics, University of Oslo, PO Box 1029
Blindern, N-0315 Oslo, Norway\\
$^5$ Astronomical Institute `Anton Pannekoek', University of Amsterdam,
Kruislaan 403, NL 1098 SJ Amsterdam, The Netherlands\\
$^6$ Dark Cosmology Centre, Niels Bohr Institute, University of
Copenhagen, DK-2100 Copenhagen, Denmark\\
$^7$ European Southern Observatory, Karl-Schwarzschild-str 2,
Garching bei M\"{u}nchen, Germany\\
}

\begin{document}
\maketitle

\begin{abstract}
We present an analysis of three strong, intervening Mg~II absorption 
systems ($z_{\rm abs} = 0.603, 0.656, 1.107$) towards the optical 
afterglow of gamma-ray burst (GRB) 060418.
From high resolution UVES spectra we measure metal
column densities and find that the highest redshift 
absorber exhibits a large 
amount of dust depletion compared with DLAs seen in QSO spectra.
The intervening $z_{\rm abs}$ = 1.107 absorber is also unusual 
in exhibiting a clear 2175 \AA\ bump, 
the first time this feature has been definitively detected in a 
GRB spectrum. The GRB afterglow spectrum is best 
fit with a two component extinction 
curve: an SMC extinction law at $z=1.49$ (the redshift of the
host) with \ebmv\ = 0.07$\pm$0.01 and a Galactic extinction curve
at $z \sim 1.1$ with \ebmv\ = 0.08$\pm$0.01.  
We also present a moderately deep NTT 
$R$-band image of the GRB060418 field and spectroscopy of four 
galaxies within 1 arcminute.  None of these objects has a redshift 
that matches any of the intervening absorbers, and we conclude
that the galaxies responsible for the two intervening Mg~II absorbers
at $z \sim 0.6$ have luminosities $\sol 0.3 L^{\star}$.

\end{abstract}

\begin{keywords}
quasars: absorption lines -- dust, extinction -- galaxies: ISM --
gamma-rays: bursts
\end{keywords}

\section{Introduction}

One of the most powerful tools in the study of the interstellar medium
(ISM) is the use of bright background sources on whose continua 
the absorption signature of intervening gas is clearly scribed. In
the Milky Way, these background sources are usually stars and for
extragalactic work, quasars.  Combined with the large apertures of
modern telescopes and the high resolving power of echelle spectrographs,
absorption line spectroscopy has yielded a detailed model of 
the ISM and the intergalactic medium (IGM) from the present day 
to z=5 (see reviews by Savage \& Sembach 1996; Rauch 1998; Wolfe, 
Gawiser \& Prochaska 2005).  Very recently, a new type of background
source has become available -- the optical afterglows of gamma-ray
bursts (GRBs).  In particular, the launch of the {\it Swift} satellite
and its rapid provision of accurate coordinates, has seen the coming 
of age of prompt, optical, ground-based follow-up of GRBs.  Although
low resolution spectra of the optical afterglow obtained a few hours after the
GRB already showed signs of intervening (as well as host galaxy)
absorption (e.g. Metzger et al. 1997; Mirabal et al. 2002; Vreeswijk et 
al. 2004), a water-shed has been recently crossed 
with the first echelle spectroscopy.  In a number of cases, this follow-up is
obtained within an hour of the burst (e.g. Chen et al. 2005), 
with the fastest follow-up being a mere 10 minutes after the GRB 
trigger (Vreeswijk et al., in preparation).  A number of exciting insights into
the GRB environment are emerging from these prompt, high resolution
data, including the detection of numerous fine structure transition lines
(Vreeswijk et al. 2004; Chen et al. 2005; Prochaska, Chen \& Bloom 2006).  

With a growing database of afterglow spectra, and with the addition
of high S/N, high resolution spectra in a number of cases, we are also
approaching the point at which preliminary statistics of {\it intervening}
absorbers can be attempted.  Prochter et al. (2006) have recently
reported a puzzling over-abundance of intervening high equivalent
width (EW) Mg~II absorbers compared with QSO sightlines.  
Three possible explanations are proposed: dust extinction in QSO
sightlines (see also Savaglio, Fall \& Fiore 2003), gravitational 
lensing in GRB sightlines and associated
(intrinsic) absorption as responsible for some of the GRB Mg~II population.
Although the first of these, bias due to dust, has been
shown to be small towards QSOs at absorption redshifts spanning 
$0.6 < z_{\rm abs} < 3.5$ (Ellison et al. 2001, 2004; Akerman et al. 
2004; Ellison, Hall \& Lira 2005; Jorgenson et al. 2006), some concerns 
about sample size remain and GRBs offer an interesting alternative
probe. 

Another attractive property of studying intervening absorbers towards
GRB afterglows is that the fading source allows for a more sensitive search
for absorber-galaxy counterparts than is possible towards QSOs.  Although
the canonical result for Mg~II absorbers is that they are
associated with relatively bright galaxies (e.g. Steidel, Dickinson
\& Persson 1994) there are
issues concerning incompleteness and mis-identification
(Churchill, Kacprzak \& Steidel 2005). In this
paper, we combine high resolution optical spectroscopy of the afterglow
to determine dust and abundance properties of the intervening
absorbers, with imaging of the GRB field and spectroscopy of low 
impact parameter galaxies. 

We use the following cosmological parameters:
$\Omega_M =0.3, \Omega_{\Lambda} = 0.7, H_0 = $70 km/s/Mpc.

\section{Data}


\begin{figure}
\centerline{\rotatebox{0}{\resizebox{8cm}{!}
{\includegraphics{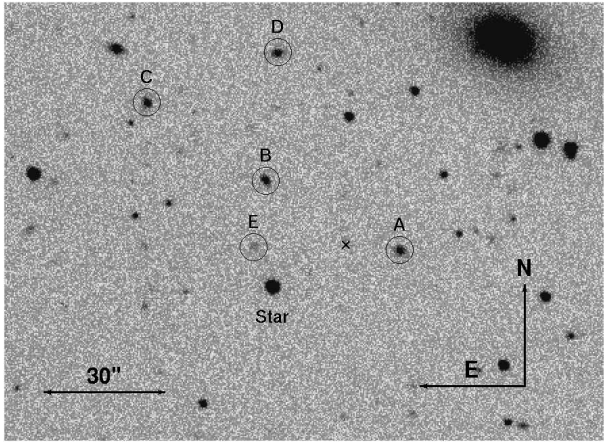}}}}
\caption{\label{grb_image} $R$-band image of the field of GRB060418
taken with EMMI on the NTT.  The objects for which we obtained spectra
are marked (A--D), as is the approximate position of the GRB (cross).
Galaxies A--D
have the following $R$-band magnitudes and redshifts (based on
absorption line matches with an elliptical galaxy template):  A --
$R$ = 21.6, $z$ = 0.22; B -- $R$ = 21.4, $z$ = 0.29; C -- $R$
= 21.1, $z$ = 0.36; D -- $R$ = 21.1, $z$ = 0.34.  Galaxy
E has a magnitude of $R$ = 22.7.}
\end{figure}

\subsection{UVES spectroscopy of the afterglow}

Our group has been involved in commisioning a Rapid Response Mode 
(RRM)\footnote{http://www.eso.org/observing/p2pp/rrm.html}
at the Very Large Telescope (VLT) in Chile.  
In this paper, we discuss the intervening absorption systems of
GRB 060418 which was observed with the RRM using the UVES echelle
spectrograph.  Details of the observations can be found in
Vreeswijk et al.(in preparation).  In brief, the wavelength
coverage is 330 -- 1000 nm, the resolution $\sim$ 7 \kms\ and the
S/N per pixel ranges from 15 -- 25.  Special care was taken
with the merging of the echelle orders, combining the coverage from 
UVES's three CCDs and flux calibration (which included a correction for
Galactic extinction with \ebmv\ = 0.224, Schlegel, Finkbeiner
\& Davis 1998).
We selected strong Mg~II absorbers with rest frame
EW(Mg~II $\lambda$ 2796) $\ge$ 1 \AA;
these are listed in Table \ref{abund_table}\footnote{Note that the
EWs quoted for Mg~II are slightly higher than the values used for the
\dind\ calculation.  In Table \ref{abund_table} we quote the full
line EW, but the \dind\ excludes the extreme velocities of the profile
where the absorption has a significance of $<3 \sigma$}.
In none of these Mg~II absorbers do we cover the \lya\ line, but  we can
apply the Ellison (2006) \dind\ as an indicator of whether the absorber
is likely to be a DLA.  In their pilot sample, Ellison (2006) found
that absorbers with
$D = 1000 \times $ EW(Mg~II $\lambda$ 2796) / $\Delta v > 6.3$ have a 90\% probability of being DLAs.
All three Mg~II systems in our sample have $D$-indices greater than 6.3.


\begin{center}
\begin{table*}
\caption{Metal Column Densities of Intervening Absorption Systems}
\begin{tabular}{lcccccccc}
\hline
$z_{\rm abs}$ & EW(Mg~II $\lambda$ 2796)$_{rest}$ & \dind\ & N(Fe~II) & N(Si~II) & N(Zn~II) & N(Cr~II) & N(Mn~II) & N(Ti~II)  \\
\hline
0.603   &  1.24$\pm$0.09 \AA  & 8.0 & 14.85$\pm$0.20 & ... & ... & ... & 12.79$\pm$0.03 & 12.51$\pm$0.03 \\ 
0.656  &1.00$\pm$0.08 \AA &  7.2 & 13.99$\pm$0.03 & ... & ... & ... & $<$12.15  &...  \\ 
1.107   & 1.87$\pm$0.06 \AA & 7.1 & 14.67$\pm$0.03 & 15.47$\pm$0.05 & 12.87$\pm$0.03 & $<$ 12.70 & 13.04$\pm$0.06  & $<$ 11.75  \\ 
\hline 
\end{tabular}\label{abund_table}
\\ All limits are 5$\sigma$.
\end{table*}
\end{center}

\subsection{Imaging}

We obtained $2 \times 300$ second exposures of the GRB field on
21 April 2006 using EMMI-RILD on the NTT at La Silla.
The final image is shown in Figure \ref{grb_image} and has a
5$\sigma$ detection limit of $R=23.45$ within a 3 arcsecond
diameter aperture.  We also obtained
spectra of four objects around the host galaxy
(galaxies A, B, C and D), with
impact parameters ranging from 13 to 60 arcseconds\footnote{1 arcsecond
is 6.69 \hkpc\ (proper) at $z=0.6$ and 8.17 \hkpc\ at $z=1.1$.}.  The spectra
were obtained with the EMMI grism \#3 yielding a FWHM resolution of
8 \AA.  The total integration time per object was 7200 seconds
for galaxies A and B and 2400 seconds for galaxies C and D.
None of the 4 objects exhibit strong emission lines but a cross-correlation
with an elliptical galaxy template yields redshifts below 0.4 (see
Figure \ref{grb_image} caption) based
on stellar absorption features.  None of these objects is therefore
likely to be associated with the strong absorption systems listed
in Table \ref{abund_table}.  The second closest object (after galaxy A)
to the GRB that is detected
in our image is galaxy E with an impact parameter of 23 arcseconds and
an approximate R band magnitude of 22.7, too faint to
follow-up with a 4-m class telescope.

\section{Relative Abundances in GRB Absorbers}


\begin{figure}
\centerline{\rotatebox{0}{\resizebox{8cm}{!}
{\includegraphics{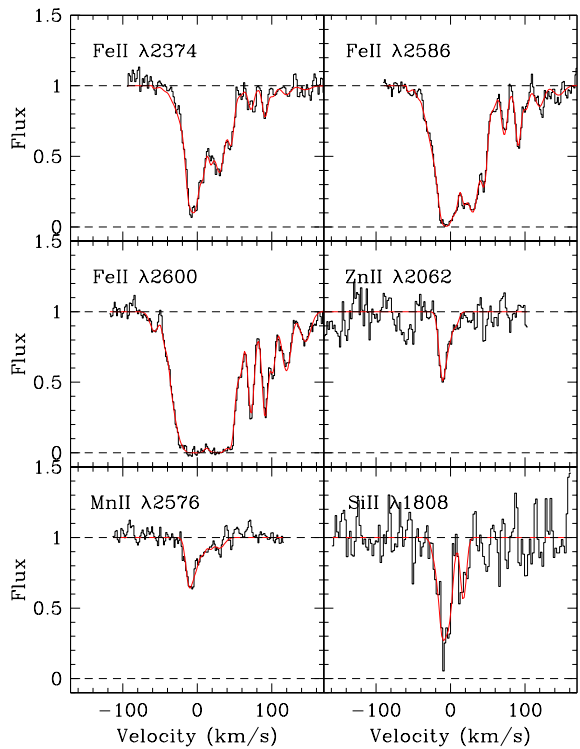}}}}
\caption{\label{fit_fig} Example Voigt profile fits for the $z_{\rm abs}$ = 1.107
absorber towards GRB060418.  }
\end{figure}

We have fitted the metal line profiles of intervening
absorbers in the UVES spectrum
with Voigt profiles using the software 
VPFIT\footnote{http://www.ast.cam.ac.uk/\~{}rfc/vpfit.html}.
The column densities are listed in Table \ref{abund_table}.
Unfortunately,
the two lower redshift systems ($z_{\rm abs} = 0.656, 0.603$)
do not have coverage of many metal lines, although the higher redshift
$z_{\rm abs} = 1.107$ has numerous transitions included
in our wavelength coverage; a selection of the absorption lines 
and fits is shown in Figure \ref{fit_fig}.  

Abundances will be driven by the combination
of underlying nucleosynthetic patterns plus the relative
susceptibility to dust depletion.  Disentangling these effects
in order to profile the chemical enrichment (and, therefore,
star formation) histories has been a long-term goal of
DLA research (e.g Dessauges-Zavadsky et al. 2006).
We do not know the absolute metallicity of the $z_{\rm abs}$ = 1.107
GRB060418 absorber due to the lack of N(HI) measurement.  However,
the large depletion factor (see below) implies that the metallicity
may be at least 1/10 of solar (Ledoux et al. 2003).
Indeed, this is the average metallicity of DLAs at $z \sim 1$
(Meiring et al. 2006), which would require a relatively high
N(HI)=21.24.   Alternatively (and taking Zn as an undepleted 
measure of metallicity),
this absorber would have solar metallicity for N(HI) = 20.24.

Even in the absence of absolute abundances, the relative
abundances that can be deduced from Table \ref{abund_table}
reveal that the  $z_{\rm abs}$ = 1.107 GRB060418 absorber
has a large amount of dust depletion (i.e. low gas-to-dust ratio).  
This is usually inferred
from the ratios of iron or chromium to zinc (e.g. Pettini 
et al. 1997), relative to the solar value: [Fe/Zn], [Cr/Zn].
We determine [Fe/Zn]=$-1.04$ which is lower than $\sim$ 95\% of DLAs
(e.g. Meiring et al. 2006)\footnote{Where necessary, we adopt 
the solar abundance scale of Lodders (2003).}.  We also measure 
[Cr/Zn] $<-1.19$ which is 
lower than all of the DLAs in the compilation of $\sim$ 40 DLAs
presented by Meiring et al. (2006).  Other abundance ratios show similar
extreme values, also indicative of dust.  For example
Si and Ti are both alpha capture elements whose Galactic stellar abundances 
track each quite well, although with a $\sim$ 0.1 dex
overabundance of Si (e.g. Fulbright 2000).  However, the
abundances of Ti and Si are very discrepant here: [Ti/Si] $<-1.10$. 
Again, this is likely due to the highly refractory nature of Ti.

\section{Extinction Curve}

\begin{figure}
\centerline{\rotatebox{270}{\resizebox{7cm}{!}
{\includegraphics{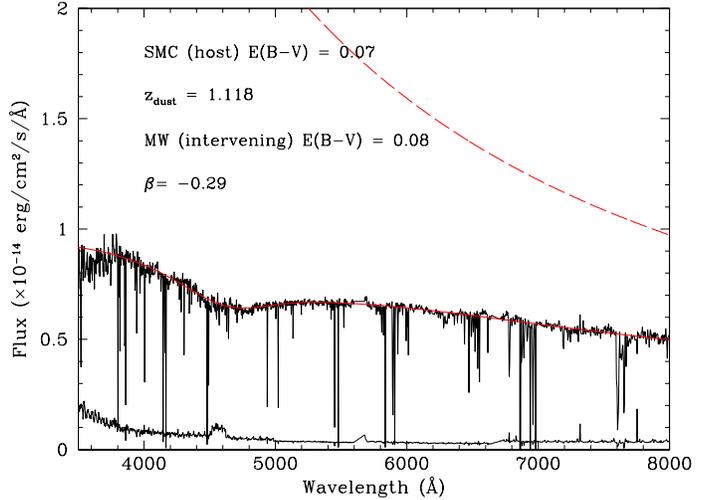}}}}
\caption{\label{ext_fig} Flux calibrated spectra and errors (black lines)
and fits (red) show the intrinsic power law spectrum (upper curve)
and dust reddened spectrum (lower curve).  The raw spectrum has been
smoothed with a Gaussian of FWHM=0.25 \AA\ for presentation purposes.
A two-component
dust fit is used with an SMC extinction curve fixed at $z=1.49$ for the
host and an intervening source of dust (redshift is fitted as
a free parameter) with a MW extinction curve.  }
\end{figure}

In order to quantify the amount of extinction that is
associated with the $z_{\rm abs} = 1.107$ absorber, we
investigate fitting the flux calibrated spectrum as a power law
continuum combined with extinction curves for the Milky
Way, SMC and LMC as parameterised by Pei (1992).  The power-law 
index ($\beta$) and  \ebmv\ are allowed to vary as free 
parameters.  We included two dust components, one fixed at the redshift 
of the host ($z=1.49$) and one for an intervening absorber
whose redshift is also a free parameter.
In the absence of a 2175 \AA\ feature, the GRB host 
galaxy extinction is best fit with an SMC curve.  However,
at $z \sim 1.1$ we find a clear 2175 \AA\ bump, indicative of carbonaceous
grains, which requires
either an LMC or MW extinction curve to achieve a good
fit.  We adopt the Galactic extinction curve as the best fit, based
on a lower $\chi^2$,  which
we show in Figure \ref{ext_fig}. The best fit parameters
for intervening Galactic-type dust are
(rest-frame, with 1$\sigma$ errors quoted) \ebmv\ = 0.08$\pm$0.01 
(A$_V$ = 0.25) at $z = 1.118^{+0.004}_{-0.001}$
and for the SMC host extinction, \ebmv\ = 0.07$\pm$0.01 (A$_V$ = 0.22).
The underlying power-law continuum has a best-fit value 
of $\beta = -0.29^{+0.04}_{-0.01}$, again with 1$\sigma$ errors
quoted based on 68\% confidence limits from the $\chi^2$ minimization.
The relatively low reddening at  $z \sim 1.1$ 
is consistent with the weak Ca~II absorption, which we measure
to have a rest frame equivalent width of only 0.1 \AA\
(Wild, Hewett \& Pettini 2006).
The extinction in the host galaxy is consistent with the
(rather wide) range of values observed in other GRB afterglow
spectra (e.g. Galama \& Wijers 2001; Vreeswijk et al. 2004).

Given the large Galactic reddening in the direction of the
GRB, we repeat the fitting process with different values of $z=0$
extinction in order to assess the sensitivity of the fit
to this parameter.  Changing the Galactic \ebmv\ by 0.1 magnitudes
results in a negligible effect on the best fit \ebmv\ of
the $z \sim 1.1$ absorber ($\Delta$ \ebmv\ = 0.003), although
the best fit $\beta$ changes by up to $\Delta \beta = 0.2$.

{\it Our data provide the first definitive detection of the 2175 \AA\
bump in
a GRB spectrum and one of the highest redshift detections of the
feature in the literature.}
The bump has only previously been observed in a handful of QSO
absorbers (e.g. Wang et al. 2004; Junkarinnen et al. 2004),
some gravitational lenses (e.g. Motta et al. 2002; Wucknitz et al.
2003) plus a tentative detection associated with the
host galaxy of GRB991216 (Vreeswijk et
al. 2006).   Reddening in QSO absorbers is usually 
best fit by an SMC-type extinction (e.g. York D. G. et al. 2006).
One interesting feature of our detection is that if the bump
occurs at $z=1.107$ (the redshift of the observed
metal lines), its inferred rest wavelength is 2186 \AA.
This is within the $\pm$ 17 \AA\ scatter seen in Galactic
sightlines, despite the theoretical prediction that more variation
should be seen for a graphite carrier (e.g. Fitzpatrick \&
Massa 1986). 


\begin{figure}
\centerline{\rotatebox{270}{\resizebox{7cm}{!}
{\includegraphics{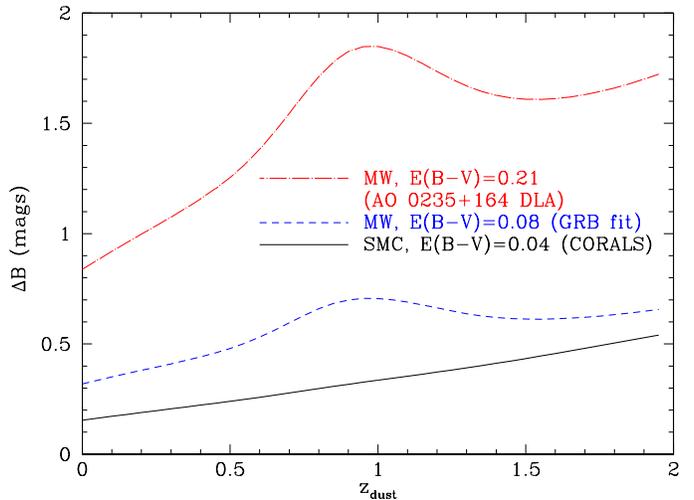}}}}
\caption{\label{db_fig} The observed frame $B$-band extinction induced
by 3 different dust models as a function of the absorber (dust)
and redshift. The dot-dash curve represents the dust present in the
$z_{\rm abs} = 0.524$ DLA towards the QSO AO 0235+164, the 
dustiest DLA known. The dashed curve illustrates extinction 
for the best fits for the
$z_{\rm abs} \sim 1.1$ absorber seen towards GRB060418.  The solid curve
represents the maximum extinction inferred from the optically complete
DLA survey of Ellison, Hall \& Lira (2005).}
\end{figure}

\section{Discussion and Conclusions}

We have presented the abundances of intervening absorbers
towards GRB060418 and, for the first time in a GRB
afterglow spectrum, detected the 2175 \AA\ bump associated with an
absorber at $z_{\rm abs}$ = 1.107. The GRB afterglow spectrum
can be well fitted with a Milky Way extinction curve
with $z_{\rm dust} \sim 1.1$ and \ebmv\ = 0.08, combined
with SMC dust at the redshift of the host.
The presence of dust in this absorber is supported by 
significantly sub-solar [Fe/Zn] and [Cr/Zn] ratios, which
are similar to the highest dust-to-gas ratios measured in QSO DLAs.  

It has been previously suggested (e.g. Savaglio et al. 2003) that
GRBs may reveal dustier absorbers than QSO surveys, since bright
afterglows could be observed even in the presence of substantial foreground
extinction.
Recently, Vladilo \& P\'eroux (2005) showed that there is a
relationship between N(Zn) and A$_V$ in the Galaxy and
in Vladilo et al. (2006) suggested a widespread correspondence
between depletion and reddening. The N(Zn) -- A$_V$ relation
applied to the $z_{\rm abs} = 1.107$ absorber towards GRB060418
predicts a rest-frame A$_V = 0.21$.  This is in good
agreement with the value determined from the Galactic extinction
curve fit (A$_V = 0.25$). As pointed out by Vladilo \& P\'eroux (2005),
this amount of extinction is quite modest, and unlikely to
produce a significant bias in absorber statistics.
We demonstrate this in Figure \ref{db_fig} where we show the
observed-frame extinction in the $B$-band of a $z = 3$ QSO
as a function of $z_{\rm dust}$.  We calculate the change
in magnitude due to extinction by taking the composite
SDSS QSO spectrum (Vanden Berk et al. 2001), applying a typical 
$B$-band transmission filter
(we use the filter curve from SUSI2 on the NTT) and applying
the extinction curves of Pei (1992). We show three curves, illustrating
the case for the $z_{\rm abs} = 1.107$ absorber towards GRB060418 and also
the $z_{\rm abs} = 0.524$ DLA towards AO 0235+164.  The latter
is one of the dustiest DLAs known with \ebmv\ = 0.23, close
to solar metallicity and even exhibiting diffuse interstellar
bands (Junkarinnen et al.2004; York, B. A. et al. 2006). In addition,
we show the effect of SMC extinction when \ebmv\ = 0.04, the
upper limit determined by Ellison et al. (2005) from \bmk\ colours
of an optically complete QSO sample.  Only the model based on
the DLA towards AO 0235+164 introduces more than $\sim 0.7$ mags of
extinction in the observed frame $B$-band.
Moreover, although the $z_{\rm abs} = 1.107$ absorber 
towards GRB060418 has extreme abundance ratios and exhibits a 2175 \AA\
bump, the observed extinction is comparable to that determined from
complete optical QSO surveys (i.e. the results of Ellison et al.
2005).  
Even with dust properties as unusual as this (i.e.
extreme depletions and a 2175 \AA\ bump), we support the 
conclusion of Prochter et al. (2006) that there is currently
no convincing evidence to suggest that dust bias is the cause
of an enhanced Mg~II number density (see also Frank et al.
2006).

Our image of the field around GRB 060418 reveals an absence
of bright galaxies ($R < 23.45$) close to the host galaxy.
The faintest galaxy within 30 arcsecs of the afterglow
(galaxy E on Figure \ref{grb_image}) would have a luminosity
of 0.7 $L^{\star}$ at $z=0.65$\footnote{We used the value
of $L^{\star}$ at $0.6 < z < 0.8$ from Ilbert et al.
(2005), including a correction to the Vega system and assuming 
that the observed frame $R$ is similar to the rest frame $B$ 
band at $z \sim 0.6$.}.  Although this is consistent
with the luminosities of Mg~II absorbers found by Steidel
et al. (1994), the impact parameter is $\sim$ 200 \hkpc,
so we conclude that this is unlikely to cause strong Mg~II
absorption in the afterglow spectrum. Our limiting magnitude 
of $R > 23.45$ corresponds to $\sim 0.3 L^{\star}$ at $z=0.65$
indicating that the absorbing
galaxies at $z \sim 0.6$ are, in this case, fainter than
many of the galaxies which have been associated with QSO
Mg~II absorbers.  For example, even when the glare of the quasar was not a 
contaminating factor, O'Meara, Chen \& Kaplan (2006)
found $\sim L^{\star}$ galaxies associated with QSO Mg~II absorbers.

In summary, based on this one field, we have found one absorber that is
apparently unusually depleted compared with QSO DLAs and is rare
in exhibiting a Galactic extinction curve with a 2175 \AA\ bump.
This is the first definitive detection of the 2175
\AA\ dust bump due to a galaxy in the line of sight to a GRB
afterglow.  We have also shown that the galaxies responsible for two
further absorbers at $z \sim 0.6$ are fainter than $0.3 L^{\star}$.
 Determining the extinction properties of more intervening
systems in GRB sightlines is required to determine how common
these properties are in galaxies seen in absorption in GRB
afterglow spectra.  The combination
of prompt and positionally accurate {\it Swift} triggers with
ground-based follow-up by teams such as our group\footnote{http://www.sc.eso.org/\~{}grbalert/uves/home.html} and GRAASP\footnote{http://graasp.ucolick.org/graasp.html} is likely to provide significant progress in this direction
in the near future.

\section*{Acknowledgments}

Based on observations made with ESO telescopes at the La Silla and Paranal
Observatories under programme ID 077.D-0661(A).
We are indebted to the staff at the VLT for their help in
commissioning and operating the RRM.  Our thanks also to
Darach Watson, Anja Andersen Hsiao-Wen Chen and Jason X. Prochaska for 
providing useful comments.

\end{document}